\documentclass[twocolumn,showpacs,preprintnumbers,amsmath,amssymb]{revtex4}

\usepackage{graphicx}
\usepackage{dcolumn}
\usepackage{bm}

\newcommand{\bfu}{\mbox{\boldmath$u$}}

\newcommand{\bfU}{\mbox{\boldmath$U$}}

\newcommand\calE{{\cal E}}

\newcommand{\bfxhat}{\mbox{\boldmath $\hat x$}}

\begin{document}


\title{Large-scale Dynamo Action Driven by Velocity Shear and Rotating Convection}

\author{David W. Hughes}
\email{d.w.hughes@leeds.ac.uk}
\affiliation{Department of Applied Mathematics, University of Leeds, LEEDS LS2 9JT, U.K.}
\author{Michael R.E. Proctor}
\email{mrep@cam.ac.uk}
\affiliation{Centre for Mathematical Sciences, University of Cambridge, Wilberforce Road, CAMBRIDGE CB3 0WA, U.K.}

\date{\today}

\begin{abstract}
By incorporating a large-scale shear flow into turbulent rotating convection, we show that a sufficiently strong shear can promote dynamo action in flows that in the absence of shear do not act as dynamos. Our results are consistent with a dynamo driven by either the shear-current effect or by the interaction between a fluctuating $\alpha$-effect and the velocity shear; they are though inconsistent with either a classical $\alpha^2$ or $\alpha \omega$ mean field dynamo.
\end{abstract}

\maketitle

Magnetic fields are observed in virtually all cosmical bodies, from planets to stars and accretion discs; in many cases their presence can be categorically attributed to dynamo action. The most pressing problem in astrophysical dynamo theory is to explain the generation of \textit{large-scale} magnetic fields; i.e.\ fields with significant energy on scales large compared with those of the driving flow. The Sun, with its global magnetic field manifested through surface activity such as sunspots, represents the most well-known example of a large-scale dynamo. 

Astrophysical dynamos are often studied within the framework of mean field electrodynamics, a tremendously elegant theory that describes the evolution of a mean (large-scale) magnetic field in terms of transport coefficients determined from averaged small-scale properties of the flow and field. The generation of magnetic field can then be ascribed to the $\alpha$-effect, which relates the mean electromotive force to the mean magnetic field. The 
$\alpha$-effect is non-zero only in flows that lack reflectional symmetry \cite{Moff}; consequently helical flows can be regarded as prime candidates for large-scale dynamo action. Indeed, in certain limiting cases the relation between $\alpha$ and helicity can be made explicit \cite{Moff,KR80}; however, and importantly, there is no theory relating these two quantities when the magnetic Reynolds number $Rm \gg 1$ and the Strouhal number $St \approx 1$, the case of astrophysical relevance. Numerical simulations reveal that the relationship between $\alpha$ and helicity is indeed far from straightforward \cite{cht06}.

The most natural system for investigating astrophysical dynamo action is that of rotating thermal convection 
\cite{cs72,sow74}. Recent studies of convection in a domain of large horizontal extent --- namely one that encompasses many convective cells --- have demonstrated that although healthy dynamo action ensues provided that $Rm$ is sufficiently large, there is no evidence of any significant large-scale magnetic field in spite of the helical nature of the convection \cite{ch06,hc08}. Indeed, attempts to measure the $\alpha$-effect directly reveal a strongly fluctuating quantity with a very small mean. The similarity of the spectra of the magnetic fields driven by rotating and non-rotating convection --- for which the flows are not helical --- provides further evidence that the dynamo is controlled by small-scale processes (such as stretching and cancellation; see \cite{CG95}) and not by mean field processes (such as a lack of reflectional symmetry).

The failure of rotating turbulent convection to act as a large-scale dynamo --- as an $\alpha^2$ dynamo in mean field parlance --- suggests that the notion that helical flows will necessarily lead to large-scale field generation is too simplistic. It is, of course, the case though that most astrophysical bodies possess a strong large-scale shear flow (differential rotation) and, indeed, most mean field astrophysical dynamo models incorporate this feature. It is therefore of interest to examine the additional effects arising from incorporating such a shear into the rotating convection model. It should though be stressed that these effects --- if favourable in terms of dynamo action --- should not disguise the inherent failure of the basic model to generate a large-scale field.

One can envisage four possible beneficial effects of the shear on the mean field dynamo process: $(i)$ that the large spatial scale of the shear leads to an enhanced $\alpha$ through greater spatial correlation of the small-scale motions \cite{hc08,cht08}; $(ii)$ that even though the mean $\alpha$ remains small there may nonetheless be an effective $\alpha \omega$ dynamo when the shear is significant; $(iii)$ that the anisotropy induced by the shear may lead to a significant shear-current effect \cite{rk03,kr07,brrk08}; $(iv)$ that the shear may interact with temporal fluctuations in $\alpha$ to produce an effective mean field dynamo \cite{Silantev,mrep07}. In this paper we describe the effects of introducing large-scale velocity shear into the model of \cite{ch06,hc08} in order to explore these various possibilities.

As in \cite{ch06,hc08} we consider a plane Boussinesq convective layer ($0 < x, y < \lambda$, $0<z<1$) with rotation about the vertical axis. In order to investigate turbulent dynamo action it is necessary to consider domains of large horizontal extent, which encompass many convective cells; this though has to be balanced by the computational demands resulting from taking $\lambda$ large. Here we take $\lambda = 5$, which results in $O(100)$ convective cells in the domain. This basic model is extended by the inclusion of a horizontal flow of the form
\begin{equation}
\bfU_0 = U_0 \cos \frac{2 \pi y}{\lambda} \bfxhat ,
\label{eq:shear}
\end{equation}
accomplished by replacing $\bfu$ with $\bfu +\bfU_0$ in the governing equations. It should be noted that although a flow with a large-scale component (i.e.\ with the same spatial dependence as (\ref{eq:shear})) does indeed result from this prescription, this component is not necessarily the `target flow' given by (\ref{eq:shear}); the hydrodynamic state that ensues depends on interactions between the shear flow and convection and, possibly, on instabilities of the shear flow itself. It is also important to note that the scale of variation of this shear flow is much greater than all scales of the convection; this is essential if the results are to be explained within the mean field framework. Tobias et al.\ \cite{tcb08} have presented results in a related geometry but with a very different shear flow, namely one that has no horizontal structure but has a strong vertical variation. 

In order to elucidate the role of shear in this initial study we focus on the regime in which convection is fairly vigorous but in which there is no dynamo action in the absence of shear; specifically we set the Rayleigh number  $Ra = 150 \, 000$, the Taylor number $Ta = 500 \, 000$, the Prandtl number $=1$ and the magnetic Prandtl number $=5$; this leads to a Reynolds number $\approx 60$ and a magnetic Reynolds number $Rm \approx 300$. A useful \textit{a priori} measure of the imposed shear is given by the shear parameter $S$, defined by
\begin{equation}
S = U_0 \left( \ell/u_{rms} L \right) ,
\label{eq:S}
\end{equation}
where $u_{rms}$ is the rms velocity in the absence of shear, $L$ is the scale of the shear and $\ell$ is the horizontal scale of the convection cells in the absence of shear. For the parameters used here, $S \approx U_0/300$. It should be pointed out that there is also an \textit{effective} value of $S$, $S_{\textrm{eff}}$ say, defined analogously to (\ref{eq:S}) but involving the shear flow that emerges dynamically in the sheared covective state; this though can only be defined \textit{a posteriori}.

\begin{figure}
\includegraphics[scale=0.41]{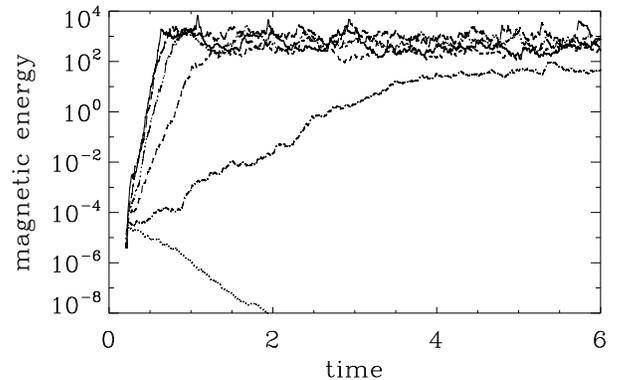}
\caption{\label{fig:1} Magnetic energy evolution for a range of $S$. In terms of increasing linear growth rate, 
$S=1/3$ (not a dynamo), $2/3$, $5/3$, $5$, $20/3$, $10/3$.}
\end{figure}

\begin{figure}
\includegraphics[scale=0.41]{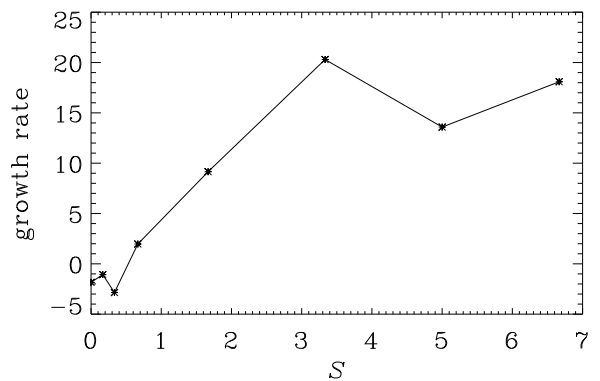}
\caption{\label{fig:2} Growth rates of the magnetic field versus $S$.}
\end{figure}

We have investigated flow and dynamo properties for the range $0 \le S \lesssim 7$. A weak seed magnetic field of zero mean is introduced into an established, stationary, purely hydrodynamic state of sheared convection. Fig.~\ref{fig:1} shows the evolution of the magnetic energy versus time for a range of values of $S$, and Fig.~\ref{fig:2} the kinematic growth rate $\gamma$ as a function of $S$. We see immediately that dynamo action ensues for sufficiently large values of $S$, although the dependence of $\gamma$ on $S$ is not straightforward. Following the onset of dynamo action (with the critical value of $S$ lying in the range $1/3 < S < 1/2$) $\gamma$ is linearly related to $S$, the strongest dependence possible \cite{Backus}. For larger $S$ though this simple relationship no longer holds. This can be explained, at least partially, by inspection of the purely hydrodynamic states. For the two largest values of $S$ considered ($S=5$, $S=20/3$), there is a transition from the mode of convection that occurs for the smaller values of $S$; specifically, the proportion of energy in the `target mode' is much smaller, leading to a reduction in $S_{\textrm{eff}}$. It is also interesting to note that for $S \gtrsim 1$ the amplitude of the saturated magnetic energy is fairly insensitive to the value of $S$.

\begin{figure}
\includegraphics[scale=0.65]{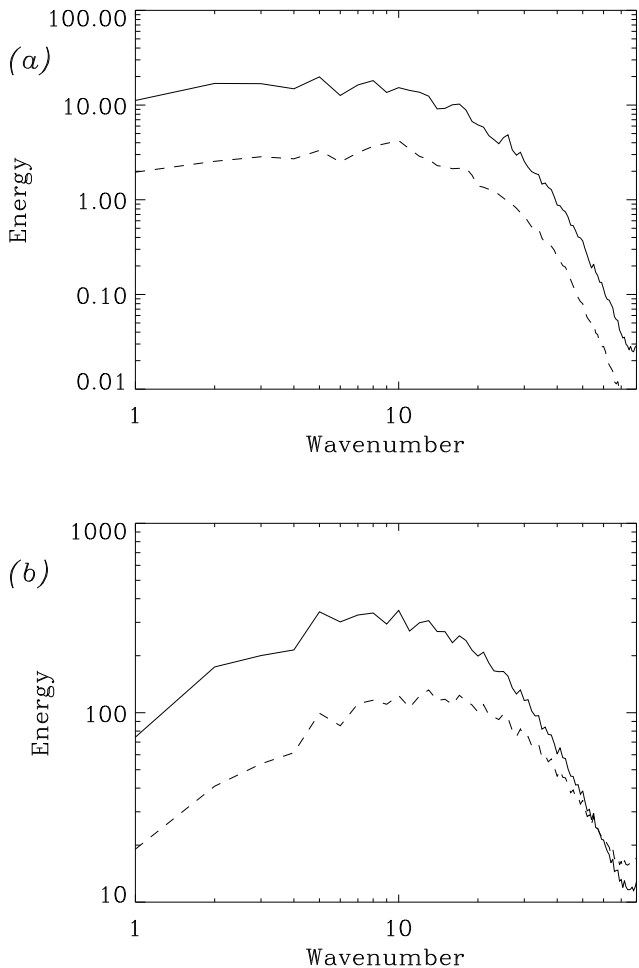}
\caption{\label{fig:3} Horizontal power spectra for the magnetic field in both the kinematic (dashed) and dynamic (solid) regimes. In $(a)$ $S = 5/3$,  $Ra = 150 \, 000$; in $(b)$ $S = 0$, $Ra = 1 \, 000 \, 000$; in both cases $Ta = 500 \, 000$. The spectra were computed over the interior region of the domain ($0.06 < z < 0.94$). The arbitrary amplitudes of the kinematic spectra have been scaled so as to be on the same plot.}
\end{figure}

\begin{figure}
\includegraphics[scale=0.6]{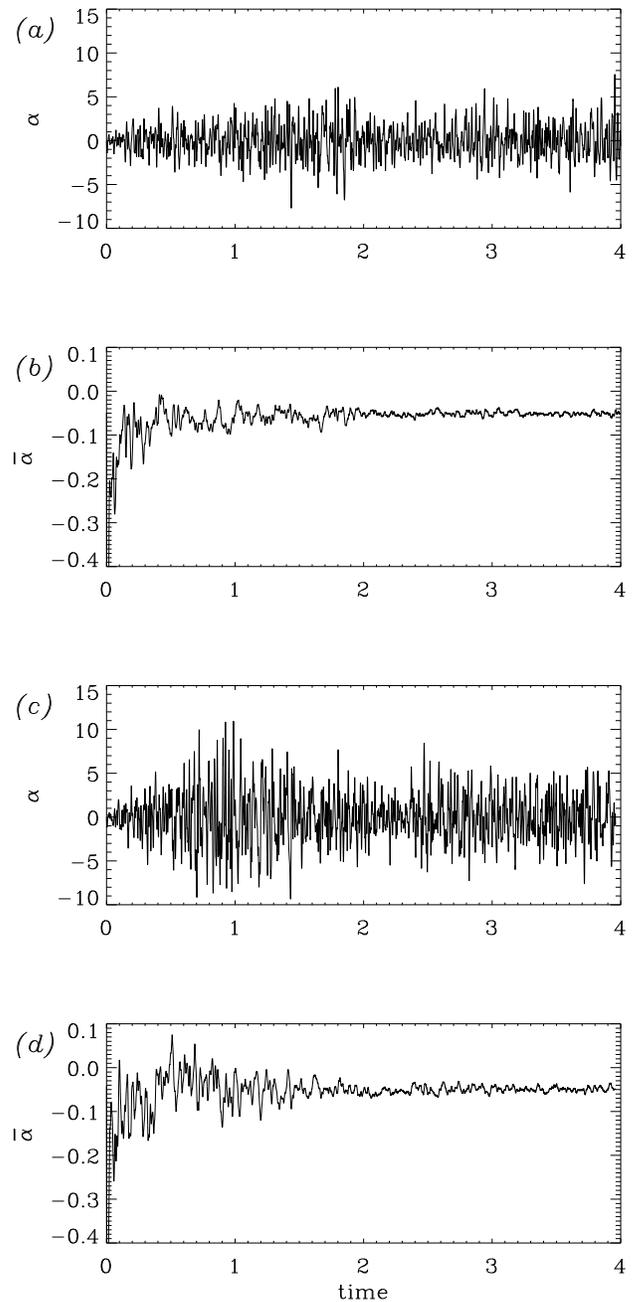}
\caption{\label{fig:4} (a) Longitudinal $\alpha$-effect versus time for $S=0$; (b) $\overline \alpha$, the cumulative temporal average of $\alpha$, for $S=0$; (c) $\alpha$ for $S=1/3$; (d) $\overline \alpha$ for $S=1/3$.}
\end{figure}

Given that our motivation was the investigation of possible large-scale dynamo action it is important to examine the spatial structure of the dynamo-generated magnetic fields, in both the kinematic and dynamic regimes. Fig.~\ref{fig:3} shows the spectra of the horizontal fields for $S=5/3$ and, for comparison, the spectra of the (small-scale) dynamo field for $Ra = 1 \, 000 \, 000$ in the absence of shear. Of particular significance is that for the case of $S= 5/3$ there is roughly equal energy in all modes of scale comparable with and greater than that of the driving convective flow, in contrast to the case of no shear, for which the spectrum is peaked at the scale of the convection. In both cases we note the similarity of the shape of the spectra in the kinematic and dynamic regimes, indicating that the structure of the field in the final nonlinear state is determined, to a large extent, from kinematic considerations.

We have also directly determined the $\alpha$-effect, by imposing a uniform horizontal magnetic field and measuring the induced electromotive force. Since this procedure has an unambiguous interpretation only in the absence of small-scale dynamo action \cite{ch08}, we have considered the value $U_0 = 100$ $(S \approx 1/3)$, which is strong enough to influence the flow but is not quite strong enough to induce dynamo action. Fig.~\ref{fig:4}$a$ shows the time history of the longitudinal $\alpha$-effect (i.e.\ $\alpha_{11}$ calculated from $\calE_x = \alpha_{11} B_{0x}$), obtained from a spatial average over half the domain \cite{ch06}, for $S=0$. As discussed in detail in \cite{hc08}, even though $\alpha$ is the result of a spatial average over many convective cells, it remains remarkably fluctuating in time, with large fluctuations and a small mean. Thus, as shown by the cumulative average in Fig.~\ref{fig:4}$b$, a further long temporal average is needed in order to pin down $\alpha$, with the resulting value being small in comparison with the rms velocity; from Fig.~\ref{fig:4}$b$ it can be seen that the long-time average value of $\alpha$ is given by $\overline{\alpha} \approx 0.05$, whereas $u_{rms} \approx 60$. For $S=1/3$, when the influence of the shear on the flow is by no means negligible, there is essentially no difference in the behaviour of $\alpha$ to that when $S=0$; as can be seen in Figs.~\ref{fig:4}$c$,$d$, it is again characterised by large fluctuations and the same small mean.

Having thus shown that the introduction of a large-scale shear flow does indeed promote vigorous large-scale dynamo action, we should return to the four possibilities discussed earlier. From the considerations immediately above, revealing $\alpha$ to be essentially unchanged by the shear flow, we can rule out possibility $(i)$. For a conventional $\alpha \omega$ dynamo model we might expect the growth rate to vary either as $S^{1/2}$ (for a disturbance of fixed wave number) or $S^{2/3}$ (if the optimal wavenumber is permitted in the system). Our calculations show that once dynamo action sets in then, for a range of $S$, the growth rate varies linearly with $S$; thus possibility $(ii)$ is also not consistent with the results. Both remaining possibilities would seem to allow the growth rate to be linearly proportional to $S$ \cite{rk07,mrep07}. However, distinguishing between the two is far from straightforward. Although the physical mechanisms are quite distinct --- the shear-current effect depends on second order velocity correlations, whereas the fluctuating $\alpha$ mechanism depends on fourth order correlations --- they are both manifested as non-diffusive contributions to the turbulent diffusivity tensor $\beta_{ijk}$.

We have demonstrated conclusively that the flow resulting from the interaction of a large-scale shear flow and turbulent rotating convection can lead to large-scale dynamo action, i.e.\ the generation of magnetic fields with a significant component of energy on scales large compared with that of the convective cells. This may be significant in understanding the generation of large-scale fields in astrophysical bodies. We have concentrated here on the regime in which the convection, although fairly vigorous, does not induce dynamo action of itself. Thus the magnetic Reynolds numbers involved are fairly modest. Obviously it is also of interest to investigate the role of shear on the small-scale dynamo action that sets in at higher $Rm$ and for which the underlying mechanism is not related to global measures, such as helicity, but instead to local stretching and folding properties of the flow, characterised, for example, by Lyapunov exponents and cancellation exponents \cite{do93}. Our results on this will be presented in a future paper.

\begin{acknowledgments}
Our work on the role of velocity shear on convective dynamos was first presented at the NORDITA programme on Turbulence and Dynamos, held in March 2008; we are grateful to the organisers for their invitation. We are also grateful to Fausto Cattaneo, who developed the code for the earlier work on convectively driven dynamos.

\end{acknowledgments}

\bibliography{Hughes_PRL}

\end{document}